\documentclass[sigconf, authorversion]{acmart}
\AtBeginDocument{%
  \providecommand\BibTeX{{%
    \normalfont B\kern-0.5em{\scshape i\kern-0.25em b}\kern-0.8em\TeX}}}

\copyrightyear{2024}
\acmYear{2024}
\setcopyright{rightsretained}
\acmConference[IDC '24]{Interaction Design and Children}{June 17--20, 2024}{Delft, Netherlands}
\acmBooktitle{Interaction Design and Children (IDC '24), June 17--20, 2024, Delft, Netherlands}
\acmDOI{10.1145/3628516.3659421}
\acmISBN{979-8-4007-0442-0/24/06}

\usepackage[caption=false]{subfig}

\begin{document}

\title{Co-designing a Child-Robot Relational Norm Intervention to Regulate Children's Handwriting Posture}

\author{Chenyang Wang}
\email{chenyang.wang@epfl.ch}
\orcid{0000-0002-9449-6953}
\affiliation{%
  \institution{CHILI Lab, EPFL}
  \streetaddress{Station 20}
  \city{Lausanne}
  \country{Switzerland}
  \postcode{1015}
}

\author{Daniel Carnieto Tozadore}
\orcid{0000-0003-0744-0132}
\affiliation{%
  \institution{CHILI Lab, EPFL}
  \streetaddress{Station 20}
  \city{Lausanne}
  \country{Switzerland}
  \postcode{1015}
}

\author{Barbara Bruno}
\orcid{0000-0003-0953-7173}
\affiliation{%
  \institution{SARAI Lab, KIT}
  \city{Karlsruhe}
  \country{Germany}
}

\author{Pierre Dillenbourg}
\orcid{0000-0001-9455-5119}
\affiliation{%
   \institution{CHILI Lab, EPFL}
   \streetaddress{Station 20}
   \city{Lausanne}
   \country{Switzerland}
   \postcode{1015}
 }
\renewcommand{\shortauthors}{Wang et al.}

\begin{abstract}
Persuasive social robots employ their social influence to modulate children's behaviours in child-robot interaction.
In this work, we introduce the Child-Robot Relational Norm Intervention (CRNI) model, leveraging the passive role of social robots and children's reluctance to inconvenience others to influence children's behaviours. 
Unlike traditional persuasive strategies that employ robots in active roles, CRNI utilizes an indirect approach by generating a disturbance for the robot in response to improper child behaviours, thereby motivating behaviour change through the avoidance of norm violations. 
The feasibility of CRNI is explored with a focus on improving children's handwriting posture.
To this end, as a preliminary work, we conducted two participatory design workshops with 12 children and 1 teacher to identify effective disturbances that can promote posture correction.
\end{abstract}



\begin{teaserfigure}
  \centering
  \includegraphics[width=0.9\textwidth]{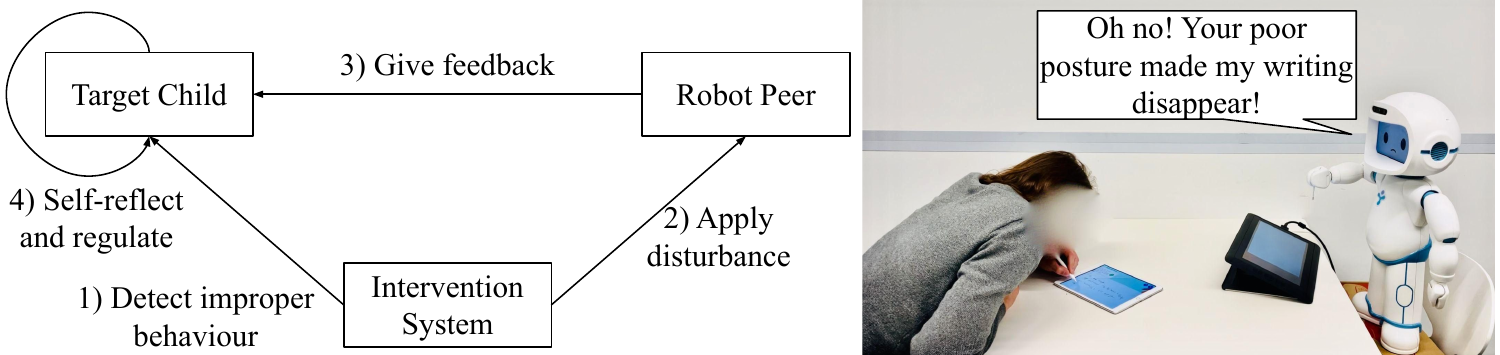}
  \caption{The proposed Child-Robot Relational Norm Intervention (CRNI)}
  \Description{This figure is split into two main sections. On the left is a flowchart describing
  the four-step Child-Robot Relational Norm Intervention (CRNI) system, and on the right is a staged photograph demonstrating the intervention in action.
The flowchart consists of four labelled steps with arrows indicating the process flow. Step 1 is "Detect improper behaviour" where the Intervention System begins the cycle by observing the target child. Step 2, "Apply disturbance," shows the intervention moving to the Robot Peer, which then influences the child. In Step 3, "Give feedback," the Robot Peer interacts with the child by providing feedback on their behaviour. Finally, in Step 4, "Self-reflect and regulate," the cycle loops back to the Target Child, suggesting an opportunity for the child to adjust their behaviour.
The photograph depicts a small humanoid robot resembling a writing on a table next to a young person slouched over a tablet on a desk. The robot is gesturing towards a speech bubble that reads, "Oh no! Your poor posture made my writing disappear!" This illustrates the feedback step where the robot points out the child's poor posture.}
  \label{fig:crni}
\end{teaserfigure}


\maketitle

\section{Introduction}
Social influence is a key strategy for persuasive technology, 
due to the fact that our behaviours are naturally and consistently influenced by the attitudes and behaviours of others depending on the social relationships we hold with them~\cite{EAGLY1984267}.
For example, people's sharing of step counts in an online social group motivates walking~\cite{StepCount2008}.
Interestingly, the power of social persuasion is not limited to human-human interaction.
Multiple studies suggest that persuasive social robots are capable of creating effective, timely and engaging persuasive human-robot interactions~\cite{liu2022systematic}.
Such persuasive social robots have been investigated in numerous application scenarios, and especially 
 for children, for example to promote healthier diets~\cite{Ilaria2014EatFruits}, encourage trash recycling~\cite{castellano2021pepperecycle}, or foster their engagement in learning activities~\cite{weiss2010robot}.

Indeed, social robots show great potential for regulating the behaviour of children. 
As the market of commercial social robots for children grows, children are increasingly interacting, engaging and forming social relationships with these robots~\cite{pashevich2022can}.
Among those interactions, compared with adults, children are more likely to perceive the robot as a social entity due to their lower awareness of the machine status of the robot~\cite{van2023transparent}, which therefore increases the chances of any social feedback provided by the robot to be taken seriously.

The current design paradigm for the persuasion strategies of social robots mainly focuses on exploiting the robot's active role to provide interventions, such as proactively giving direct reminders or commands from a role designed to convey authority, like teacher or manager, or suggestions from peer roles~\cite{saunderson2021persuasive, blancas2015effects}.
Conversely, in this work, we propose a persuasion strategy employing the passive role of social robots, by using children's reluctance to inconvenience others~\cite{Stout2001OtherRegardingPreferences}, which is named \textit{Child-Robot Relational Norm Intervention (CRNI)} in reference to the model of Relational Norm Intervention~\cite{Shin2016BeUpright} by Shin~\textit{et al.} which inspires it.

The fundamental principle of CRNI is the assumption that leveraging relational norms between children and robots has the potential to shape their behaviour.
As shown in Figure~\ref{fig:crni}, CRNI can be used to regulate certain behaviours of children, such as their sitting posture while handwriting on a tablet.
Whenever an improper behaviour, such as a wrong sitting posture, is detected, the system will not directly intervene towards the child, but rather generate an aversive stimulus against the robot as a disturbance. 
As a response to the aversive stimulus, the robot will complain and thus ``passively ask'' the child to correct their behaviour.
In this way, the CRNI model creates an artificial causal relationship between the improper behaviour of the child and the robot's disturbance.
More formally, the CRNI model relies on children’s tendency to avoid norm violations and postulates using it as a motivation for behaviour change due to Other-Regarding Preferences~\cite{Stout2001OtherRegardingPreferences}.

To preliminarily investigate the feasibility of the CRNI approach we apply it to the case of handwriting posture regulation. The motivation for this choice is the sadly pervasive and frequent poor posture taken by children during writing, which can have long-lasting negative consequences~\cite{quka2015risk}.
In line with this choice, we specifically apply our CRNI paradigm in the context of the iReCHeCk framework~\cite{irecheck2022,irecheck}, a Child-Robot Interaction system for handwriting learning support envisioning a social robot to interact with and guide children in practising handwriting on tablets.

We argue that the design of the disturbance applied to the robot is crucial for the effectiveness of the CRNI approach.
On the one hand, it should be ``powerful enough'' to dissuade children from breaching relational norms and evoke empathy for the robot in a way that motivates them to change their behaviours. On the other hand, it should be ``mild enough'' to not elicit emotional distress in the children \cite{rosenthal2014investigations}.
To identify ``the right medium'', 
in this work we conducted two consecutive participatory design workshops on disturbance design with 12 children and 1 teacher from a local primary school, in which we specifically investigated the following two research questions:

\textit{RQ1}: What disturbance(s) against robot peers do children perceive as more effective in the context of CRNI for handwriting posture regulation?


\textit{RQ2}: Are children's proposed disturbances against robot peers different from those they suggest for human peers?

\section{Regulating Children’s Posture When Interacting with Tablets via CRNI}
\label{sec:interaction_context}
Maintaining a good posture during the interactions with tablets is vital for children, as poor posture can lead to physical health problems~\cite{quka2015risk}, and proper posture can enhance learning performance, particularly in tasks that involve motor skills like handwriting~\cite{rosenblum2006relationships, Wang:311523}.
Multiple methods have been investigated to regulate the body posture of children when interacting with tablets, including dedicated garments constraining the body movement~\cite{Alsuwaidi2018}, automated direct notifications~\cite{bootsman2019wearable} and unobtrusive regulation by moving UI elements on the tablet screen~\cite{wang2024visualstimuli,Wang2024WriteUpRight}.
In this work, we try to employ the power of the avoidance of child-robot social norm violation, as captured by the proposed CRNI model.

\textit{Interaction Context}: 
Located in the same classroom as the child, the robot plays the role of a peer for the child. Both are asked to engage in handwriting practice at the same time.
Following the paradigm of Figure~\ref{fig:crni}, whenever the child adopts an inappropriate posture, the system will disturb the robot peer, prompting the robot to complain and give feedback to the child, thus making the child aware of the consequences of their poor posture.
We postulate that this method can stimulate self-reflection in children and prompt more self-regulation and even lasting behaviour change~\cite{lee2011altruistic}.
Being a robot already used for children's handwriting learning support~\cite{irecheck2022}, QTrobot\footnote{https://luxai.com/} (shown in Figure~\ref{fig:crni}) is adopted in the proposed CRNI model as the robot peer.
We argue that its cartoon-style appearance and other dedicated designs like the child-like voice might facilitate the creation of an emotional and relational bond by children~\cite{van2020child}.

\section{Participatory Design}
Two workshops were conducted with the same group of children. 
Each lasted approx. 45 minutes.
In the first workshop, we presented the CRNI idea to the children and brainstormed with them different designs for the disturbance to be applied to the robot peer.
The most voted ideas were then implemented and showcased in a live demo of the CRNI approach for handwriting posture regulation in the second workshop, where we collected further feedback.

12 children from a local international primary school were invited to participate in the study by contacting the teacher and upon consent from their parents\footnote{This study has received ethical approval from the Human Research Ethics Committee of EPFL under protocol HREC 057-2021. Informed consent was given in written form by the parents, prior to the beginning of the experiment, and orally by the children at the beginning of each session.}.
All children belong to the same class of Year 5  (4 girls and 8 boys aged $M$ = 9.42 years old, $SD$ = 0.51).
The children come from diverse cultural and socioeconomic backgrounds and all use spoken and written English in their daily lives at school.
Digital tablets are frequently used for writing in class activities.
Their class teacher was involved and present throughout the study.

\subsection{Design Workshop I}
\begin{figure*}[!t]
    \centering
    \begin{subfloat}[Phase 1 storytelling (left) and Phase 3 robot self-introduction (right) of Workshop I]{
    \includegraphics[width=0.46\linewidth]{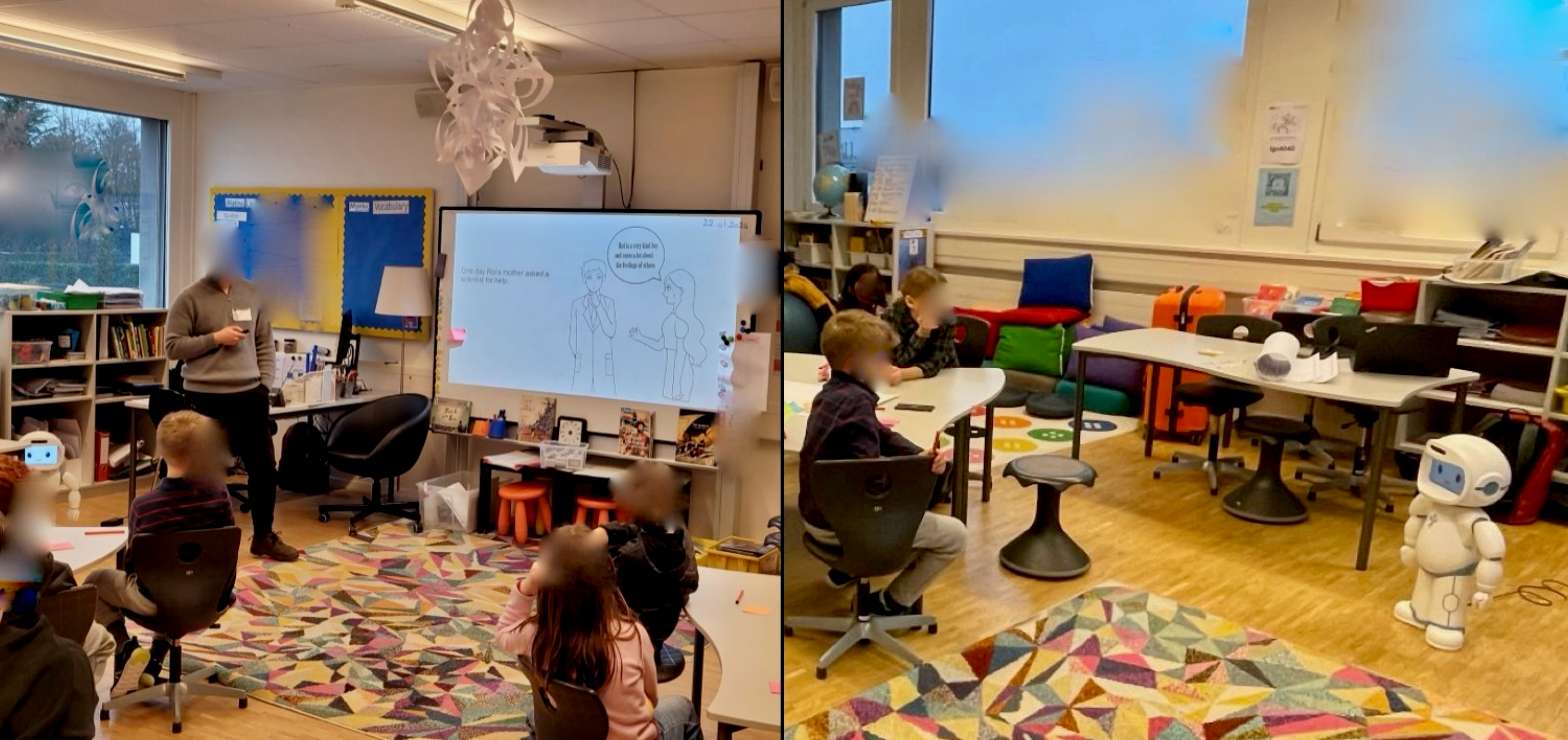}
    \label{fig:classroom_robot_intro}}
    \Description{This figure showcases two photographs side by side, labelled as Phase 1 storytelling (left) and Phase 3 robot self-introduction (right) of Workshop I. 
In the left photo, we observe an indoor classroom setting where children are seated on chairs, attentively facing a large screen displaying a drawing of two characters with a dialogue bubble. A host stands next to the screen, facing towards the children, engaging in storytelling.
The right photo presents a different angle of what appears to be the same room, now with the focus on a humanoid robot positioned to the right of the frame. The robot, which has a screen for a face and a predominantly white body, stands near a table. The robot's arms are raised slightly, suggesting an interactive or welcoming posture as if it is introducing itself. A group of children, seen from the back, are seated and facing the robot with their attention directed towards it.
}
    \end{subfloat}
    \begin{subfloat}[The lower part of the poster B that is used for the disturbance design for robot peers]{
    \includegraphics[width=0.46\linewidth]{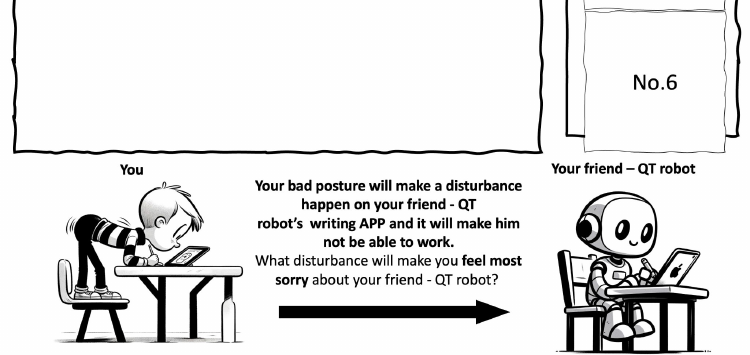}
    \label{fig:poster_2}}
    \Description{
This figure, labeled as (b) The lower part of the poster B, depicts a conceptual design for a disturbance-based intervention involving a child and a QT robot, intended to address and correct the child's posture.
On the left side of the image, a boy labeled "You" hunched over a tablet on a desk, demonstrating poor posture. To the right, labelled "Your friend – QT robot," the QT robot is seated upright at a desk, attentively looking at a tablet. Above the image is a narrative suggesting a cause and effect relationship: "Your bad posture will make a disturbance happen on your friend - QT robot's writing APP and it will make him not be able to work. What disturbance will make you feel most sorry about your friend - QT robot?" This text implies an interactive element where the child's posture affects the robot's ability to operate, thus encouraging the child to reflect on their posture and its impact on the robot.
    }
   \end{subfloat}
    \caption{Photos and poster B of Workshop I
    }
    \label{fig:interaction-design}
\end{figure*}
\subsubsection{Method and Materials}\hspace*{1mm}
\label{subsec:method_workshop1}
The collaborative ideation method known as ``sticky note brainstorming''~\cite{christensen2020properties} was employed in the first workshop.
The children were divided into 3 groups of 4 children by their teacher, each group seated at separate tables in the same classroom and 3 hosts from the research team coordinated the activity of the groups.
Each group was provided with two A3-sized posters (poster A and B) as illustrated in Figures~\ref{fig:poster_2} and ~\ref{fig:result-group-a}. 
%
The sticky note brainstorming was designed as follows: Firstly, each child was asked to think independently and write their ideas on sticky notes of their distinct colour, then they presented their ideas within the group and placed their notes on the left blank area of the poster. 
Subsequently, three disturbances were proposed to the children by a researcher: 1) arbitrarily scaling the reading text field of the peer’s writing App, 2) locking the tablet of the peer’s writing App, and 3) emptying the ink of the peer's digital pencil.
The disturbances were intentionally introduced after the brainstorming session to avoid influencing children's thinking.
Following this, each host grouped similar sticky notes and stacked them together. 
Children in the same group could then vote on the aggregated ideas, and the sorted results would be organized on the ranking board to the right by the host.
Each child was asked to vote via stickers for the two ideas they considered most effective.
In the end, the host reviewed the rankings and summarized the outcomes for the children.

\subsubsection{Procedures}\hspace*{1mm}
To facilitate children's understanding of the CRNI model and its use with QTrobot for handwriting posture regulation, and to prompt more creative designs, we organised the workshop in the following four phases:

\textit{Phase 1: Design Context Storytelling.}
The workshop started with a storytelling introduction given by a researcher as shown in Figure~\ref{fig:classroom_robot_intro}, in which an original comic story about \textit{Roi and His Bad Posture}
\footnote{
A short description of the story is as follows: Roi is an 8-year-old boy who enjoys handwriting but often does it with a poor posture, leading to occasional neck pain. Concerned about his well-being, Roi's mother seeks help from a scientist. Recognizing Roi's kindness and sensitivity towards others, the scientist devises an intelligent device to aid him. This device, resembling a camera, monitors Roi's posture during handwriting tasks and controls two iPads used by Roi and his friend Lily during class. If Roi adopts a bad posture, the device disrupts Lily's iPad by locking it for two minutes, causing her frustration. Lily confronts Roi about the issue, prompting him to apologize and correct his posture. Through this experience, Roi learns the importance of maintaining a good posture while writing. 
Available at \url{https://doi.org/10.5281/zenodo.10845067}} was presented.
The story was designed to illustrate, simply and clearly, the application context and the principle of relational norm regulation~\cite{Shin2016BeUpright}.
The comic narrative incorporated the example disturbance ``lock the friend's iPad' to elucidate the concept and facilitate children's comprehension of the disturbances.

\textit{Phase 2: Design for Human Peer.}
To further facilitate children's comprehension of the task, the storytelling was followed by a session of sticky notes for the design of disturbances applied to human peers (a scenario closer to the story and to children's daily experience), working on poster A in Figure~\ref{fig:result-group-a}.
We introduced the design objective by letting the \textit{scientist} of the comic story ask questions: \textit{``What if other children do not care about locking their friends’ iPads? What other disturbances can we design?''}.
The session was conducted as outlined in Section~\ref{subsec:method_workshop1}.

\textit{Phase 3: Design for Robot Peer.}
At the beginning of the design session focusing on disturbances for robot peers (RQ1), QTrobot introduced itself in the role of an 8-year-old child from Luxembourg as shown in Figure~\ref{fig:classroom_robot_intro}. The introduction lasted approx. 5 minutes.
The design context of QTrobot as a peer was then illustrated in the presentation and emphasized in the new poster B as shown in Figure~\ref{fig:poster_2}.
Each host transferred all sticky notes from poster A to the blank section of poster B, then encouraged the children to come up with new ideas, which were then presented within the group and deliberated upon as described in Section~\ref{subsec:method_workshop1}. The difference between the outcome of this phase with respect to the previous one allows for investigating RQ2.
%
%

\textit{Phase 4: Preference Poll.}
At the end of the workshop, we inquired with the children about their preference between engaging in the activity with a peer, as in Phase 2, or with QTrobot, as in Phase 3. 
The children were instructed to indicate their preference by first raising their hands for the human peer option and subsequently for QTrobot.

\subsection{Design Workshop II}
To further support the children's understanding of CRNI and their creativity, 
we implemented the top three disturbance designs for QTrobot (see Figure~\ref{fig:rank-robot-peer}) that emerged from Workshop I and showcased them during Workshop II, which took place around 7 weeks later.
We argue that introducing tangible examples of disturbances applied to the robot within the participatory design process not only motivates and engages the children but also fosters and channels their creative thinking skills, leading to more meaningful and impactful design contributions.

\subsubsection{Method and Materials}\hspace*{1mm}
The QTrobot and a Wacom Cintiq display~\cite{url_wacom}
were teleoperated by a researcher, while another acted as the child writing in a bad posture.
This solution enables the QTrobot to simulate letter writing in the air, while the corresponding visuals appear on the Wacom display in real-time. 
We prepared a worksheet\footnote{Available at \url{https://doi.org/10.5281/zenodo.10845067}} for each child which allows them to assess the disturbances with guidance by ranking them before and after the demonstration.
In addition, the worksheet allows them to give textual feedback about the justification for the rankings.

\subsubsection{Procedures}\hspace*{1mm}
Unlike Workshop I, Workshop II envisioned children to complete their worksheets independently.
The teacher was also invited to participate in the activity and work on the worksheet.
We started the session with a brief recap of Workshop I's findings, then invited the children to rank the three shortlisted disturbances on the worksheet based on their perceived effectiveness.
This activity was followed by a 6-minute demonstration of the CRNI model with QTrobot.
During the demo, a researcher played the role of the target child by performing bad body posture while writing on the iPad and another researcher operated the QTrobot to timely and sequentially initiate the three disturbances and the robot's reaction to them,
while the children gathered around the QTrobot to observe its experience and reactions.
After the demonstration, the children were asked to rank the disturbances again and justify their rankings on the worksheet.
The workshop ended with a reward session where children played games with QTrobot, followed by a 10-minute interview with the teacher.

\begin{figure}[!t]
    \centering
    \begin{subfloat}[Design results for the human friend in Workshop I]{
    \includegraphics[width=0.95\linewidth]{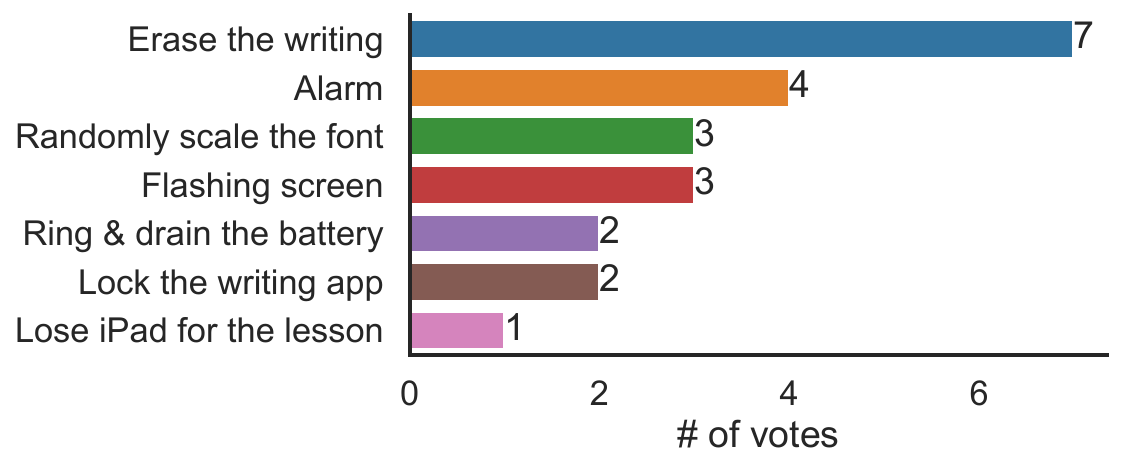}
    \label{fig:rank-human-peer}}
    \Description{
    This figure presents a horizontal bar chart with the title (a) Design results for the human friend in Workshop I. The chart depicts the number of votes received for different types of interventions that could be applied to the child's friend.
Each bar represents an intervention type, with the length of the bar corresponding to the number of votes it received from workshop participants. Starting from the top, the most voted intervention, with 7 votes, is "Erase the writing," shown by a blue bar. The second is "Alarm" with 4 votes, represented by an orange bar. "Randomly scale the font" and "Flashing screen" each have 3 votes and are indicated by green and red bars respectively. Following are "Ring & drain the battery" and "Lock the writing app" with 2 votes each, shown by light blue and brown bars. Lastly, "Lose iPad for the lesson" has the least votes, with 1, depicted by a pink bar.
}
    \end{subfloat}
    \begin{subfloat}[Design results for QTrobot in Workshop I]{
    \includegraphics[width=0.95\linewidth]{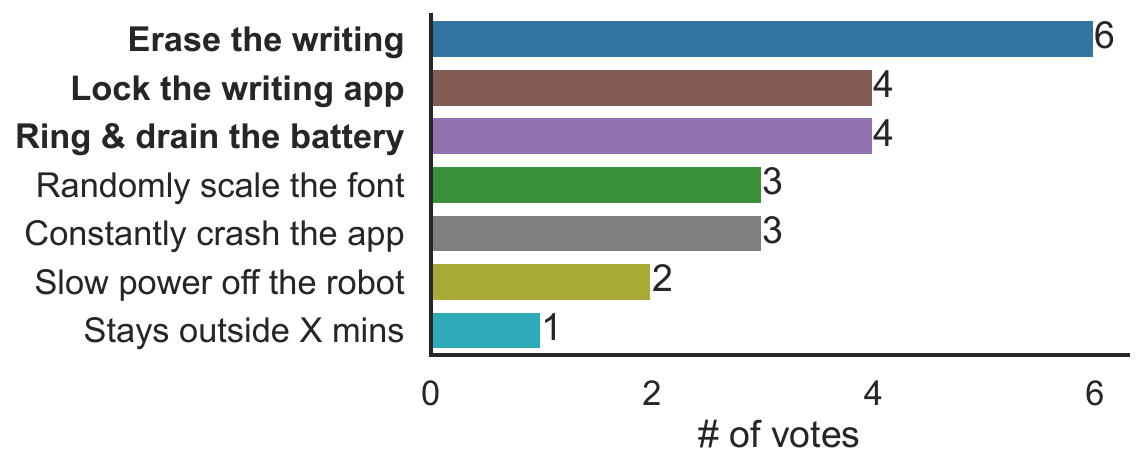}
    \label{fig:rank-robot-peer}}
    \Description{
    This bar chart is titled (b) Design results for QTrobot in Workshop I and illustrates the number of votes for different disturbance interventions designed for the QTrobot, as determined by the workshop participants.
The bars are colour-coded to differentiate between the types of interventions. The intervention "Erase the writing" received the most votes, shown by a blue bar, with a total of 6. Both "Lock the writing app" and "Ring & drain the battery" interventions are next, each with 4 votes, indicated by brown and purple bars respectively. "Randomly scale the font" and "Constantly crash the app" interventions follow with 3 votes each, shown with green and grey bars. The "Slow power off the robot" has 2 votes, represented by a light green bar. The least voted intervention, with 1 vote, is "Stays outside X mins," displayed with a light blue bar.
    }
   \end{subfloat}
   \vspace{-5pt} 
    \caption{Design results of Workshop I. The top three bold font designs were showcased in Workshop II.
    }
    \label{fig:workshop1-design}
\end{figure}

\section{Results And Discussion}
\label{sec:result}

\subsection{Children's Designs in Workshop I}

In Workshop I, each of the 3 groups produced two lists of sorted disturbance designs as Figure~\ref{fig:result-group-a}: one for a human friend and another for QTrobot.
We summarized the result by grouping all the voted designs together.
The top 7 designs for each context by votes are illustrated in Figure~\ref{fig:workshop1-design}.
As the Figure shows, the children believe that the most effective disturbance 
is to ``erase the writing homework'', for both types of peers. 

The comparison of the two lists allows for investigating the different design choices made by children when considering disturbances for human or robot peers (RQ2) and provides interesting insights into children's perception of the robot.
Options such as  ``alarm'' and ``flashing screen'', which children considered quite effective for human peers, disappear from the list for QTrobot, while designs like ``ring \& drain the battery of tablet'' and ``lock the writing app'' received more votes for a robot peer than a human peer.
Furthermore, new designs specifically tailored for a robot were proposed by children during the robot disturbances design phase, like ``slowly power off the robot'' and ``let the robot stay outside for X minutes''. 
Most of the top disturbances for the robot peer seem to exclusively aim to affect its performance in the task, thus suggesting that children viewed the robot as a cognitive agent.
The disturbance of letting the robot stay outside might indicate a certain degree of anthropomorphization of the robot, and empathy for it on the side of the children.
This further implies the potential persuasive power of the CRNI model.

\subsection{Children's Design Assessment in Workshop II}
\label{subsec:workshop2}
\begin{figure}[!tbh]
    \centering
    \begin{subfloat}[]{
    \includegraphics[width=0.95\linewidth]{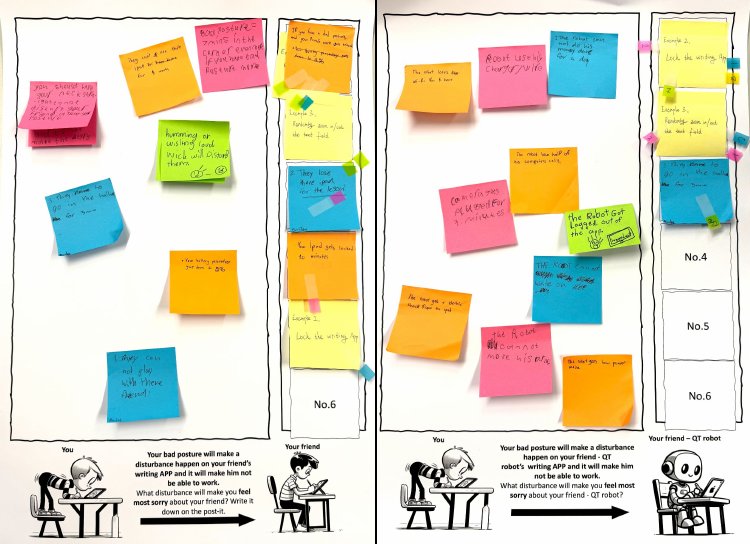}
    \label{fig:result-group-a}}
    \Description{
    This figure displays two posters filled with colourful sticky notes, each representing the disturbance design results from a group in Workshop I. The left poster (Poster A), is intended for human peers, and on the right poster (Poster B), is designed for robot peers.
}
    \end{subfloat}
    \begin{subfloat}[]{
    \includegraphics[width=0.95\linewidth]{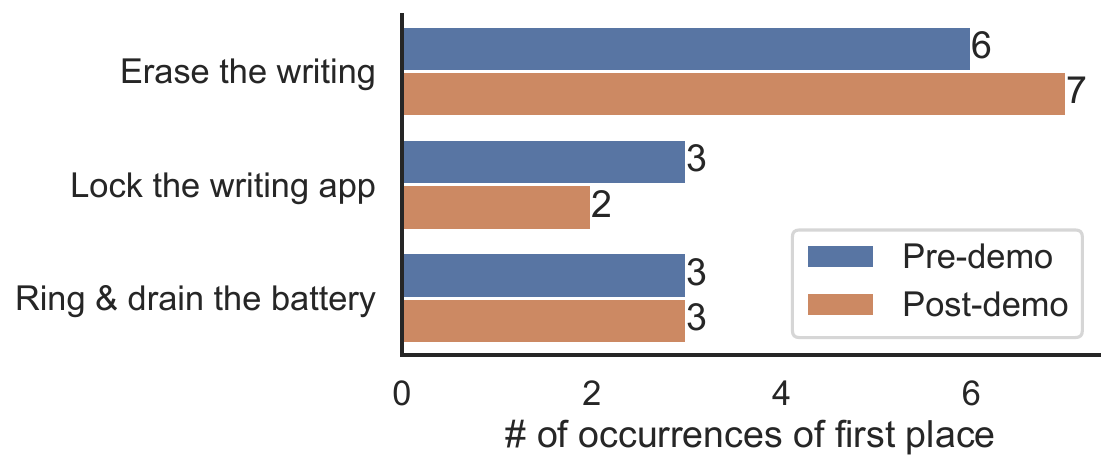}
    \label{fig:workshop2-result}}
    \Description{
    The figure is a bar chart that indicates the final design assessment of children in Workshop II. The chart lists three interventions: "Erase the writing," "Lock the writing app," and "Ring & drain the battery." Two sets of bars represent the number of occurrences of each intervention being ranked in the first place before (blue bars) and after (orange bars) a demonstration (pre-demo and post-demo). "Erase the writing" was the most frequently selected as the top intervention in both pre-demo (6 occurrences) and post-demo (7 occurrences). "Lock the writing app" had 3 occurrences pre-demo and 2 post-demo, while "Ring & drain the battery" had equal occurrences (3) in both assessments.
    }
   \end{subfloat}
    \caption{ (a) The disturbance design results of one group in Workshop I. Left poster A for human peers and right poster B for robot peers. (b) The final design assessment of children in Workshop II.
    }
    \label{fig:poster-and-workshop2-result}
\end{figure}

Based on the collected worksheets, we counted the number of times each implemented disturbance was the top-rated one, pre- and post-demonstration (see Figure~\ref{fig:workshop2-result}).
``Erase the writing'', the top pick in Workshop I, was also the preferred choice before the demonstration in Workshop II, as well as after it.
It is worth noticing that two girls verbally stated ``I feel so bad when seeing the QT's writing is gone!'' during the demonstration, and 
ample other empathetic responses were textually provided by the children.
One child 
wrote: ``QT is writing really hard on something then it gets deleted. I would feel so bad about it.''
Likewise, another child commented: ``I think deleting the homework is bad because QT does not know how to write and it will be hard for him to do it again! It would be very sad if all your work would be gone.''
Therefore, we conclude that ``erasing the writing'' might be a suitable candidate as the disturbance design for a CRNI model for handwriting posture regulation (RQ1).

An interesting gender effect was noticed in the post-demonstration rankings, where all 4 girls independently picked ``erasing the writing'' as the most effective disturbance.
Conversely, the 8 boys had more diverse rankings.
Since the literature on gender bias in empathy is inconclusive, with some studies suggesting the existence of gender differences~\cite{benenson2021girls} and some not~\cite{pang2023women}, we will investigate eventual gender differences in the efficacy of the CRNI model in future studies.


\subsection{Teacher's Feedback}
The teacher participated in Workshop II. 
In line with the children's design assessment (Figure~\ref{fig:workshop2-result}), the teacher also picked ``erasing the writing'' as the disturbance capable of eliciting the strongest empathetic response, both pre- and post-demonstration.
During the interview, 
she expressed reservations about the ability of children in the age of her class to anthropomorphize the robot, 
which stands in contrast to the children's empathetic behaviours as reported in Section~\ref{subsec:workshop2}.
She also expressed concerns about the growing reluctance of children for handwriting exercises, particularly when contrasted with the keyboard typing training at school.

\subsection{Robot Peer vs Human Peer}
According to children's vote in the preference poll phase, the majority of children (10 out of 12) stated they prefer using QTrobot in the given activity rather than a human friend. 
Children's reasons for this choice included feeling more motivated and supported by the robot. 
This aligns with research in Child-Robot Interaction suggesting social robots can motivate learning~\cite{weiss2010robot} and behaviour change~\cite{deshmukh2019influencing}, although they might be biased by the novelty effect. 
In this regard, the integration of a social robot in handwriting practice could be a strong motivator to alleviate children's resistance towards it, thus helping tackle the concerns expressed by the teacher.
Additionally, the use of a robot peer within CRNI can provide consistent availability and alleviate ethical concerns compared to using a human peer.


\subsection{Limitations and Future Works}
While the workshops provided insightful feedback for our future experiments, it's worth acknowledging some limitations.
First, while we focused on the design of disturbances applied to the robot peer, other elements of the CRNI model deserve equal attention, such as the optional warning to be sent to the child prior to triggering the disturbance and the duration of the disturbance itself.
Furthermore, our co-design activities directly asked children about the disturbances that they thought would most likely elicit the reluctance of norm violation.
Consequently, the children's designs and feedback were rooted in their subjective and hypothetical empathy towards the robot peer, which may not align with the actual reaction in real-life situations.
This potential discrepancy will be investigated in future research, together with the level of distress experienced by children upon witnessing the robot being disturbed.
Lastly, the regulatory efficacy of the CRNI model in real-world settings remains untested.
Future studies should particularly examine whether the CRNI model can produce immediate, lasting, or even habit-forming changes in children's handwriting posture.

\begin{acks}
We would like to express our gratitude to the teacher and all the students who participated in the study. 
We are also grateful to Mr. Zhenyu Cai and Ms. Jie Gao for their invaluable assistance in our research.
\end{acks}

\bibliographystyle{ACM-Reference-Format}
\bibliography{mybib}


\begin{thebibliography}{29}


\ifx \showCODEN    \undefined \def \showCODEN     #1{\unskip}     \fi
\ifx \showDOI      \undefined \def \showDOI       #1{#1}\fi
\ifx \showISBNx    \undefined \def \showISBNx     #1{\unskip}     \fi
\ifx \showISBNxiii \undefined \def \showISBNxiii  #1{\unskip}     \fi
\ifx \showISSN     \undefined \def \showISSN      #1{\unskip}     \fi
\ifx \showLCCN     \undefined \def \showLCCN      #1{\unskip}     \fi
\ifx \shownote     \undefined \def \shownote      #1{#1}          \fi
\ifx \showarticletitle \undefined \def \showarticletitle #1{#1}   \fi
\ifx \showURL      \undefined \def \showURL       {\relax}        \fi
\providecommand\bibfield[2]{#2}
\providecommand\bibinfo[2]{#2}
\providecommand\natexlab[1]{#1}
\providecommand\showeprint[2][]{arXiv:#2}

\bibitem[Alsuwaidi et~al\mbox{.}(2018)]%
        {Alsuwaidi2018}
\bibfield{author}{\bibinfo{person}{Alyazyah Alsuwaidi}, \bibinfo{person}{Aisha Alzarouni}, \bibinfo{person}{Dana Bazazeh}, \bibinfo{person}{Nawaf Almoosa}, \bibinfo{person}{Kinda Khalaf}, {and} \bibinfo{person}{Raed Shubair}.} \bibinfo{year}{2018}\natexlab{}.
\newblock \bibinfo{title}{Wearable Posture Monitoring System with Vibration Feedback}.
\newblock
\newblock
\showeprint[arxiv]{1810.00189}~[eess.SP]


\bibitem[Baroni et~al\mbox{.}(2014)]%
        {Ilaria2014EatFruits}
\bibfield{author}{\bibinfo{person}{Ilaria Baroni}, \bibinfo{person}{Marco Nalin}, \bibinfo{person}{Mattia Coti~Zelati}, \bibinfo{person}{Elettra Oleari}, {and} \bibinfo{person}{Alberto Sanna}.} \bibinfo{year}{2014}\natexlab{}.
\newblock \showarticletitle{Designing motivational robot: How robots might motivate children to eat fruits and vegetables}. In \bibinfo{booktitle}{\emph{The 23rd IEEE International Symposium on Robot and Human Interactive Communication}}. \bibinfo{publisher}{IEEE}, \bibinfo{address}{Edinburgh, UK}, \bibinfo{pages}{796--801}.
\newblock
\urldef\tempurl%
\url{https://doi.org/10.1109/ROMAN.2014.6926350}
\showDOI{\tempurl}


\bibitem[Benenson et~al\mbox{.}(2021)]%
        {benenson2021girls}
\bibfield{author}{\bibinfo{person}{Joyce~F Benenson}, \bibinfo{person}{Evelyne Gauthier}, {and} \bibinfo{person}{Henry Markovits}.} \bibinfo{year}{2021}\natexlab{}.
\newblock \showarticletitle{Girls exhibit greater empathy than boys following a minor accident}.
\newblock \bibinfo{journal}{\emph{Scientific Reports}} \bibinfo{volume}{11}, \bibinfo{number}{1} (\bibinfo{year}{2021}), \bibinfo{pages}{7965}.
\newblock


\bibitem[Blancas et~al\mbox{.}(2015)]%
        {blancas2015effects}
\bibfield{author}{\bibinfo{person}{Maria Blancas}, \bibinfo{person}{Vasiliki Vouloutsi}, \bibinfo{person}{Klaudia Grechuta}, {and} \bibinfo{person}{Paul~FMJ Verschure}.} \bibinfo{year}{2015}\natexlab{}.
\newblock \showarticletitle{Effects of the robot’s role on human-robot interaction in an educational scenario}. In \bibinfo{booktitle}{\emph{Biomimetic and Biohybrid Systems: 4th International Conference, Living Machines 2015, Barcelona, Spain, July 28-31, 2015, Proceedings 4}}. Springer, \bibinfo{pages}{391--402}.
\newblock


\bibitem[Bootsman et~al\mbox{.}(2019)]%
        {bootsman2019wearable}
\bibfield{author}{\bibinfo{person}{Rik Bootsman}, \bibinfo{person}{Panos Markopoulos}, \bibinfo{person}{Qi Qi}, \bibinfo{person}{Qi Wang}, {and} \bibinfo{person}{Annick~AA Timmermans}.} \bibinfo{year}{2019}\natexlab{}.
\newblock \showarticletitle{Wearable technology for posture monitoring at the workplace}.
\newblock \bibinfo{journal}{\emph{International Journal of Human-Computer Studies}}  \bibinfo{volume}{132} (\bibinfo{year}{2019}), \bibinfo{pages}{99--111}.
\newblock


\bibitem[Castellano et~al\mbox{.}(2021)]%
        {castellano2021pepperecycle}
\bibfield{author}{\bibinfo{person}{Giovanna Castellano}, \bibinfo{person}{Berardina De~Carolis}, \bibinfo{person}{Francesca D’Errico}, \bibinfo{person}{Nicola Macchiarulo}, {and} \bibinfo{person}{Veronica Rossano}.} \bibinfo{year}{2021}\natexlab{}.
\newblock \showarticletitle{PeppeRecycle: Improving children’s attitude toward recycling by playing with a social robot}.
\newblock \bibinfo{journal}{\emph{International Journal of Social Robotics}}  \bibinfo{volume}{13} (\bibinfo{year}{2021}), \bibinfo{pages}{97--111}.
\newblock


\bibitem[Chenyang et~al\mbox{.}(2024)]%
        {Wang2024WriteUpRight}
\bibfield{author}{\bibinfo{person}{Wang Chenyang}, \bibinfo{person}{Daniel Carnieto~Tozadore}, \bibinfo{person}{Barbara Bruno}, {and} \bibinfo{person}{Pierre Dillenbourg}.} \bibinfo{year}{2024}\natexlab{}.
\newblock \showarticletitle{WriteUpRight: Regulating Children’s Handwriting Body Posture by Unobstrusively Error Amplification via Slow Visual Stimuli on Tablets}.
\newblock \bibinfo{journal}{\emph{Proceedings of the CHI Conference on Human Factors in Computing Systems (CHI '24)}} (\bibinfo{year}{2024}).
\newblock
\urldef\tempurl%
\url{https://doi.org/10.1145/3613904.3642457}
\showDOI{\tempurl}


\bibitem[Christensen et~al\mbox{.}(2020)]%
        {christensen2020properties}
\bibfield{author}{\bibinfo{person}{Bo~T Christensen}, \bibinfo{person}{Kim Halskov}, {and} \bibinfo{person}{Clemens~N Klokmose}.} \bibinfo{year}{2020}\natexlab{}.
\newblock \showarticletitle{The properties of sticky notes for collaborative creativity: An introduction}.
\newblock In \bibinfo{booktitle}{\emph{Sticky Creativity}}. \bibinfo{publisher}{Elsevier}, \bibinfo{pages}{1--16}.
\newblock


\bibitem[Consolvo et~al\mbox{.}(2008)]%
        {StepCount2008}
\bibfield{author}{\bibinfo{person}{Sunny Consolvo}, \bibinfo{person}{David~W. McDonald}, \bibinfo{person}{Tammy Toscos}, \bibinfo{person}{Mike~Y. Chen}, \bibinfo{person}{Jon Froehlich}, \bibinfo{person}{Beverly Harrison}, \bibinfo{person}{Predrag Klasnja}, \bibinfo{person}{Anthony LaMarca}, \bibinfo{person}{Louis LeGrand}, \bibinfo{person}{Ryan Libby}, \bibinfo{person}{Ian Smith}, {and} \bibinfo{person}{James~A. Landay}.} \bibinfo{year}{2008}\natexlab{}.
\newblock \showarticletitle{Activity sensing in the wild: a field trial of ubifit garden}. In \bibinfo{booktitle}{\emph{Proceedings of the SIGCHI Conference on Human Factors in Computing Systems}} (Florence, Italy) \emph{(\bibinfo{series}{CHI '08})}. \bibinfo{publisher}{Association for Computing Machinery}, \bibinfo{address}{New York, NY, USA}, \bibinfo{pages}{1797–1806}.
\newblock
\showISBNx{9781605580111}
\urldef\tempurl%
\url{https://doi.org/10.1145/1357054.1357335}
\showDOI{\tempurl}


\bibitem[Deshmukh et~al\mbox{.}(2019)]%
        {deshmukh2019influencing}
\bibfield{author}{\bibinfo{person}{Amol Deshmukh}, \bibinfo{person}{Sooraj~K Babu}, \bibinfo{person}{R Unnikrishnan}, \bibinfo{person}{Shanker Ramesh}, \bibinfo{person}{Parameswari Anitha}, {and} \bibinfo{person}{Rao~R Bhavani}.} \bibinfo{year}{2019}\natexlab{}.
\newblock \showarticletitle{Influencing hand-washing behaviour with a social robot: Hri study with school children in rural india}. In \bibinfo{booktitle}{\emph{2019 28th IEEE international conference on robot and human interactive communication (RO-MAN)}}. IEEE, \bibinfo{publisher}{IEEE}, \bibinfo{address}{New Delhi, India}, \bibinfo{pages}{1--6}.
\newblock


\bibitem[Eagly and Chaiken(1984)]%
        {EAGLY1984267}
\bibfield{author}{\bibinfo{person}{Alice~H. Eagly} {and} \bibinfo{person}{Shelly Chaiken}.} \bibinfo{year}{1984}\natexlab{}.
\newblock \showarticletitle{Cognitive Theories of Persuasion}.
\newblock In \bibinfo{booktitle}{\emph{Advances in Experimental Social Psychology}}, \bibfield{editor}{\bibinfo{person}{Leonard Berkowitz}} (Ed.). \bibinfo{series}{Advances in Experimental Social Psychology}, Vol.~\bibinfo{volume}{17}. \bibinfo{publisher}{Academic Press}, \bibinfo{pages}{267--359}.
\newblock
\showISSN{0065-2601}
\urldef\tempurl%
\url{https://doi.org/10.1016/S0065-2601(08)60122-7}
\showDOI{\tempurl}


\bibitem[Lee et~al\mbox{.}(2011)]%
        {lee2011altruistic}
\bibfield{author}{\bibinfo{person}{Yeoreum Lee}, \bibinfo{person}{Youn-kyung Lim}, {and} \bibinfo{person}{Hyeon-Jeong Suk}.} \bibinfo{year}{2011}\natexlab{}.
\newblock \showarticletitle{Altruistic interaction design: a new interaction design approach for making people care more about others}. In \bibinfo{booktitle}{\emph{Proceedings of the 2011 Conference on Designing Pleasurable Products and Interfaces}} (Milano, Italy) \emph{(\bibinfo{series}{DPPI '11})}. \bibinfo{publisher}{Association for Computing Machinery}, \bibinfo{address}{New York, NY, USA}, Article \bibinfo{articleno}{9}, \bibinfo{numpages}{4}~pages.
\newblock
\showISBNx{9781450312806}
\urldef\tempurl%
\url{https://doi.org/10.1145/2347504.2347514}
\showDOI{\tempurl}


\bibitem[Liu et~al\mbox{.}(2022)]%
        {liu2022systematic}
\bibfield{author}{\bibinfo{person}{Baisong Liu}, \bibinfo{person}{Daniel Tetteroo}, {and} \bibinfo{person}{Panos Markopoulos}.} \bibinfo{year}{2022}\natexlab{}.
\newblock \showarticletitle{A systematic review of experimental work on persuasive social robots}.
\newblock \bibinfo{journal}{\emph{International Journal of Social Robotics}} \bibinfo{volume}{14}, \bibinfo{number}{6} (\bibinfo{year}{2022}), \bibinfo{pages}{1339--1378}.
\newblock


\bibitem[Pang et~al\mbox{.}(2023)]%
        {pang2023women}
\bibfield{author}{\bibinfo{person}{Chenyu Pang}, \bibinfo{person}{Wenxin Li}, \bibinfo{person}{Yuqing Zhou}, \bibinfo{person}{Tianyu Gao}, {and} \bibinfo{person}{Shihui Han}.} \bibinfo{year}{2023}\natexlab{}.
\newblock \showarticletitle{Are women more empathetic than men? Questionnaire and EEG estimations of sex/gender differences in empathic ability}.
\newblock \bibinfo{journal}{\emph{Social cognitive and affective neuroscience}} \bibinfo{volume}{18}, \bibinfo{number}{1} (\bibinfo{year}{2023}), \bibinfo{pages}{nsad008}.
\newblock


\bibitem[Pashevich(2022)]%
        {pashevich2022can}
\bibfield{author}{\bibinfo{person}{Ekaterina Pashevich}.} \bibinfo{year}{2022}\natexlab{}.
\newblock \showarticletitle{Can communication with social robots influence how children develop empathy? Best-evidence synthesis}.
\newblock \bibinfo{journal}{\emph{AI \& SOCIETY}} \bibinfo{volume}{37}, \bibinfo{number}{2} (\bibinfo{year}{2022}), \bibinfo{pages}{579--589}.
\newblock


\bibitem[Quka et~al\mbox{.}(2015)]%
        {quka2015risk}
\bibfield{author}{\bibinfo{person}{Najada Quka}, \bibinfo{person}{DH Stratoberdha}, {and} \bibinfo{person}{R Selenica}.} \bibinfo{year}{2015}\natexlab{}.
\newblock \showarticletitle{Risk factors of poor posture in children and its prevalence}.
\newblock \bibinfo{journal}{\emph{Academic Journal of Interdisciplinary Studies}} \bibinfo{volume}{4}, \bibinfo{number}{3} (\bibinfo{year}{2015}), \bibinfo{pages}{97}.
\newblock


\bibitem[Rosenblum et~al\mbox{.}(2006)]%
        {rosenblum2006relationships}
\bibfield{author}{\bibinfo{person}{Sara Rosenblum}, \bibinfo{person}{Sarina Goldstand}, {and} \bibinfo{person}{Shula Parush}.} \bibinfo{year}{2006}\natexlab{}.
\newblock \showarticletitle{Relationships among biomechanical ergonomic factors, handwriting product quality, handwriting efficiency, and computerized handwriting process measures in children with and without handwriting difficulties}.
\newblock \bibinfo{journal}{\emph{The American journal of occupational therapy}} \bibinfo{volume}{60}, \bibinfo{number}{1} (\bibinfo{year}{2006}), \bibinfo{pages}{28--39}.
\newblock


\bibitem[Rosenthal-Von Der~P{\"u}tten et~al\mbox{.}(2014)]%
        {rosenthal2014investigations}
\bibfield{author}{\bibinfo{person}{Astrid~M Rosenthal-Von Der~P{\"u}tten}, \bibinfo{person}{Frank~P Schulte}, \bibinfo{person}{Sabrina~C Eimler}, \bibinfo{person}{Sabrina Sobieraj}, \bibinfo{person}{Laura Hoffmann}, \bibinfo{person}{Stefan Maderwald}, \bibinfo{person}{Matthias Brand}, {and} \bibinfo{person}{Nicole~C Kr{\"a}mer}.} \bibinfo{year}{2014}\natexlab{}.
\newblock \showarticletitle{Investigations on empathy towards humans and robots using fMRI}.
\newblock \bibinfo{journal}{\emph{Computers in Human Behavior}}  \bibinfo{volume}{33} (\bibinfo{year}{2014}), \bibinfo{pages}{201--212}.
\newblock


\bibitem[Saunderson and Nejat(2021)]%
        {saunderson2021persuasive}
\bibfield{author}{\bibinfo{person}{Shane~P Saunderson} {and} \bibinfo{person}{Goldie Nejat}.} \bibinfo{year}{2021}\natexlab{}.
\newblock \showarticletitle{Persuasive robots should avoid authority: The effects of formal and real authority on persuasion in human-robot interaction}.
\newblock \bibinfo{journal}{\emph{Science robotics}} \bibinfo{volume}{6}, \bibinfo{number}{58} (\bibinfo{year}{2021}), \bibinfo{pages}{eabd5186}.
\newblock


\bibitem[Shin et~al\mbox{.}(2016)]%
        {Shin2016BeUpright}
\bibfield{author}{\bibinfo{person}{Jaemyung Shin}, \bibinfo{person}{Bumsoo Kang}, \bibinfo{person}{Taiwoo Park}, \bibinfo{person}{Jina Huh-Yoo}, \bibinfo{person}{Jinhan Kim}, {and} \bibinfo{person}{Junehwa Song}.} \bibinfo{year}{2016}\natexlab{}.
\newblock \showarticletitle{BeUpright: Posture Correction Using Relational Norm Intervention}.
\newblock \bibinfo{journal}{\emph{Proceedings of the SIGCHI conference on human factors in computing systems CHI Conference}}  \bibinfo{volume}{2016}, \bibinfo{pages}{6040--6052}.
\newblock
\urldef\tempurl%
\url{https://doi.org/10.1145/2858036.2858561}
\showDOI{\tempurl}


\bibitem[Stout(2001)]%
        {Stout2001OtherRegardingPreferences}
\bibfield{author}{\bibinfo{person}{Lynn Stout}.} \bibinfo{year}{2001}\natexlab{}.
\newblock \showarticletitle{Other-Regarding Preferences and Social Norms}.
\newblock \bibinfo{journal}{\emph{SSRN Electronic Journal}} (\bibinfo{date}{03} \bibinfo{year}{2001}).
\newblock
\urldef\tempurl%
\url{https://doi.org/10.2139/ssrn.265902}
\showDOI{\tempurl}


\bibitem[Tozadore et~al\mbox{.}(2023)]%
        {irecheck}
\bibfield{author}{\bibinfo{person}{Daniel~C. Tozadore}, \bibinfo{person}{Soizic Gauthier}, \bibinfo{person}{Barbara Bruno}, \bibinfo{person}{Chenyang Wang}, \bibinfo{person}{Jianling Zou}, \bibinfo{person}{Lise Aubin}, \bibinfo{person}{Dominique Archambault}, \bibinfo{person}{Mohamed Chetouani}, \bibinfo{person}{Pierre Dillenbourg}, \bibinfo{person}{David Cohen}, {and} \bibinfo{person}{Salvatore~M. Anzalone}.} \bibinfo{year}{2023}\natexlab{}.
\newblock \showarticletitle{The iReCheck project: using tablets and robots for personalised handwriting practice}. In \bibinfo{booktitle}{\emph{Companion Publication of the 25th International Conference on Multimodal Interaction}} (Paris, France) \emph{(\bibinfo{series}{ICMI '23 Companion})}. \bibinfo{publisher}{Association for Computing Machinery}, \bibinfo{address}{New York, NY, USA}, \bibinfo{pages}{297–301}.
\newblock
\showISBNx{9798400703218}
\urldef\tempurl%
\url{https://doi.org/10.1145/3610661.3616178}
\showDOI{\tempurl}


\bibitem[Tozadore et~al\mbox{.}(2022)]%
        {irecheck2022}
\bibfield{author}{\bibinfo{person}{Daniel~Carnieto Tozadore}, \bibinfo{person}{Chenyang Wang}, \bibinfo{person}{Giorgia Marchesi}, \bibinfo{person}{Barbara Bruno}, {and} \bibinfo{person}{Pierre Dillenbourg}.} \bibinfo{year}{2022}\natexlab{}.
\newblock \showarticletitle{A game-based approach for evaluating and customizing handwriting training using an autonomous social robot}. In \bibinfo{booktitle}{\emph{2022 31st IEEE International Conference on Robot and Human Interactive Communication (RO-MAN)}}. \bibinfo{publisher}{IEEE}, \bibinfo{address}{Napoli, Italy}, \bibinfo{pages}{1467--1473}.
\newblock
\urldef\tempurl%
\url{https://doi.org/10.1109/RO-MAN53752.2022.9900661}
\showDOI{\tempurl}


\bibitem[van Straten et~al\mbox{.}(2020)]%
        {van2020child}
\bibfield{author}{\bibinfo{person}{Caroline~L van Straten}, \bibinfo{person}{Jochen Peter}, {and} \bibinfo{person}{Rinaldo K{\"u}hne}.} \bibinfo{year}{2020}\natexlab{}.
\newblock \showarticletitle{Child--robot relationship formation: A narrative review of empirical research}.
\newblock \bibinfo{journal}{\emph{International Journal of Social Robotics}} \bibinfo{volume}{12}, \bibinfo{number}{2} (\bibinfo{year}{2020}), \bibinfo{pages}{325--344}.
\newblock


\bibitem[van Straten et~al\mbox{.}(2023)]%
        {van2023transparent}
\bibfield{author}{\bibinfo{person}{Caroline~L van Straten}, \bibinfo{person}{Jochen Peter}, {and} \bibinfo{person}{Rinaldo K{\"u}hne}.} \bibinfo{year}{2023}\natexlab{}.
\newblock \showarticletitle{Transparent robots: How children perceive and relate to a social robot that acknowledges its lack of human psychological capacities and machine status}.
\newblock \bibinfo{journal}{\emph{International Journal of Human-Computer Studies}}  \bibinfo{volume}{177} (\bibinfo{year}{2023}), \bibinfo{pages}{103063}.
\newblock


\bibitem[Wacom(2024)]%
        {url_wacom}
\bibfield{author}{\bibinfo{person}{Wacom}.} \bibinfo{year}{2024}\natexlab{}.
\newblock \bibinfo{booktitle}{\emph{Wacom Cintiq}}.
\newblock Wacom.
\newblock
\urldef\tempurl%
\url{https://www.wacom.com/en-ch/products/pen-displays/wacom-cintiq}
\showURL{%
\tempurl}


\bibitem[Wang et~al\mbox{.}(2024)]%
        {Wang:311523}
\bibfield{author}{\bibinfo{person}{Chenyang Wang}, \bibinfo{person}{Daniel Carnieto~Tozadore}, \bibinfo{person}{Barbara Bruno}, {and} \bibinfo{person}{Pierre Dillenbourg}.} \bibinfo{year}{2024}\natexlab{}.
\newblock \showarticletitle{Can Children Benefit from Technological Applications for Body Posture Correction to Improve Handwriting? A Study to Quantitatively Investigate the Correlation between Body Posture and Handwriting Quality}.
\newblock  (\bibinfo{year}{2024}).
\newblock
\urldef\tempurl%
\url{https://doi.org/10.5075/epfl-labo-311523}
\showDOI{\tempurl}


\bibitem[Wang et~al\mbox{.}(2023)]%
        {wang2024visualstimuli}
\bibfield{author}{\bibinfo{person}{Chenyang Wang}, \bibinfo{person}{Daniel~C. Tozadore}, \bibinfo{person}{Barbara Bruno}, {and} \bibinfo{person}{Pierre Dillenbourg}.} \bibinfo{year}{2023}\natexlab{}.
\newblock \showarticletitle{Unobtrusively Regulating Children’s Posture via Slow Visual Stimuli on Tablets}. In \bibinfo{booktitle}{\emph{Proceedings of the 22nd Annual ACM Interaction Design and Children Conference}} (Chicago, IL, USA) \emph{(\bibinfo{series}{IDC '23})}. \bibinfo{publisher}{Association for Computing Machinery}, \bibinfo{address}{New York, NY, USA}, \bibinfo{pages}{548–552}.
\newblock
\showISBNx{9798400701313}
\urldef\tempurl%
\url{https://doi.org/10.1145/3585088.3593894}
\showDOI{\tempurl}


\bibitem[Weiss et~al\mbox{.}(2010)]%
        {weiss2010robot}
\bibfield{author}{\bibinfo{person}{A Weiss}, \bibinfo{person}{T Scherndl}, \bibinfo{person}{R Buchner}, {and} \bibinfo{person}{M Tscheligi}.} \bibinfo{year}{2010}\natexlab{}.
\newblock \showarticletitle{A robot as persuasive social actor a field trial on child-robot interaction}. In \bibinfo{booktitle}{\emph{Proceedings of the 2nd international symposium on new frontiers in human--robot interaction—a symposium at the AISB 2010 convention}}. \bibinfo{pages}{136--142}.
\newblock


\end{thebibliography}










\end{document}